 \documentclass[aps,prx,twocolumn,showpacs,superscriptaddress,longbibliography]{revtex4-2}
\usepackage{graphicx}
\usepackage{epstopdf}
\usepackage{float}
\usepackage{color}
\usepackage{amsmath}
\usepackage{bm}

\begin{document}

\title{$^{55}$Mn-NMR/NQR study on the kagome metal Sc$_3$Mn$_3$Al$_7$Si$_5$: Re-examining the possibility of spin-singlet paired state}

\author{Qing-Ping Ding}
\affiliation{Ames National Laboratory, U.S. DOE, Ames, Iowa 50011, USA}

\author{Yongbin Lee}
\affiliation{Ames National Laboratory, U.S. DOE, Ames, Iowa 50011, USA}

\author{Charuni Dissanayake}
\email{Present address: National High Magnetic Field Laboratory, Florida State University, Tallahassee, FL 32310}
\affiliation{Department of Physics, University of Central Florida, Orlando, Florida 32816, USA}




\author{Yasuyuki Nakajima}
\affiliation{Department of Physics, University of Central Florida, Orlando, Florida 32816, USA}

\author{Yuji Furukawa}
\email{ furukawa@ameslab.gov}
\affiliation{Ames National Laboratory, U.S. DOE, Ames, Iowa 50011, USA}
\affiliation{Department of Physics and Astronomy, Iowa State University, Ames, Iowa 50011, USA}

\date{\today}

\begin{abstract}
     
     Sc$_3$Mn$_3$Al$_7$Si$_5$ is a Mn-based kagome metal that has been proposed to be close to a ferromagnetic instability. Recent $^{55}$Mn nuclear magnetic resonance (NMR) study suggested the emergence of a spin-singlet paired state below $\sim$10 K. To clarify its low-energy magnetic and electronic properties, we performed $^{55}$Mn NMR and nuclear quadrupole resonance (NQR)  measurements on single-crystalline Sc$_3$Mn$_3$Al$_7$Si$_5$.
     From the detailed NMR/NQR  spectrum measurements and the DFT calculations, a quadrupolar frequency of $\nu_{\rm Q}$ = 4.6$\sim$4.8 MHz and an asymmetry parameter of $\eta$ = 0.45$\sim$0.48 as well as the principal axes of electric field gradient EFG for the Mn site were determined.  
     The temperature dependence of  Knight shift ($K$) shows a similar temperature dependence of the DC  magnetic susceptibility $\chi$ except for the low-temperature region below $\sim$ 50 K where  $K$ is almost constant while $\chi$ keeps increasing, which suggests that the increase in $\chi$ at low temperatures is not intrinsic.
     $^{55}$Mn spin-lattice relaxation rate divided by temperature ($1/T_1T$)  is found to be nearly constant,  confirming the metallic state of Sc$_3$Mn$_3$Al$_7$Si$_5$ from a microscopic point of view. 
     Furthermore, no peak in $1/T_1T$ is observed near 10 K, and neither the Knight shift nor the linewidth shows any reductions at low temperatures. 
     These results suggest no clear signature for the spin-singlet pairing around 10 K in Sc$_3$Mn$_3$Al$_7$Si$_5$.

  \end{abstract}

\maketitle
 \section{Introduction}
 
    Magnetic fluctuations are increasingly recognized as a key ingredient in understanding the exotic properties of kagome metals. 
   Owing to the coexistence of geometric frustration, flat electronic bands, and nontrivial band topology, kagome systems are highly sensitive to electronic and magnetic instabilities. 
   In recent years, magnetic fluctuations have been implicated in a variety of unusual phenomena, including the large topological Hall effect in YMn$_6$Sn$_6$ \cite{Ghimire2020} and the non-Fermi-liquid behavior and superconductivity in LaRu$_3$Si$_2$ \cite{Wang2023}. 
    Clarifying the nature of magnetic fluctuations is, therefore, essential for understanding the microscopic properties of kagome materials.

   A wide variety of correlated and topological states has been discovered in kagome compounds      
  $RM_6$Sn$_6$ family ($R$ = rare-earth element and $M$ = 3$d$ transition metal), often referred to as the 166-kagome series,  is one of relatively well-studied kagome metals \cite{Xu2023}. 
    In ferrimagnetic TbMn$_6$Sn$_6$, spin-polarized Dirac fermions, anomalous Hall effects, and Chern topological magnetism have been reported \cite{Yin2020,Riberolles2022,Riberolles2023,Riberolles2024}. 
    Furthermore, nearly dispersionless edge states have been predicted within the Dirac gap when spin-orbit coupling and out-of-plane ferromagnetism are taken into account \cite{Riberolles2024}. 
    $R$V$_6$Sn$_6$ system also shows rich ground states, including charge-density-wave order in ScV$_6$Sn$_6$ \cite{Cao2023}, antiferromagnetism in YbV$_6$Sn$_6$ \cite{Guo2023,Park2026}, and ferromagnetism in GdV$_6$Sn$_6$ \cite{Pokharel2021} and TbV$_6$Sn$_6$ \cite{Pokharel2022}. 
   Similar complexity is observed in the $A$V$_3$Sb$_5$ ($A$ = K, Rb, and Cs) family, where superconductivity coexists or competes with charge-density waves, and electronic nematicity \cite{Ortiz2019,Ortiz2020,Wilson2024}. 
  These findings make kagome metals as fertile systems for exploring such emergent quantum phenomena.

 \begin{figure}[b]
\includegraphics[width=0.9\columnwidth]{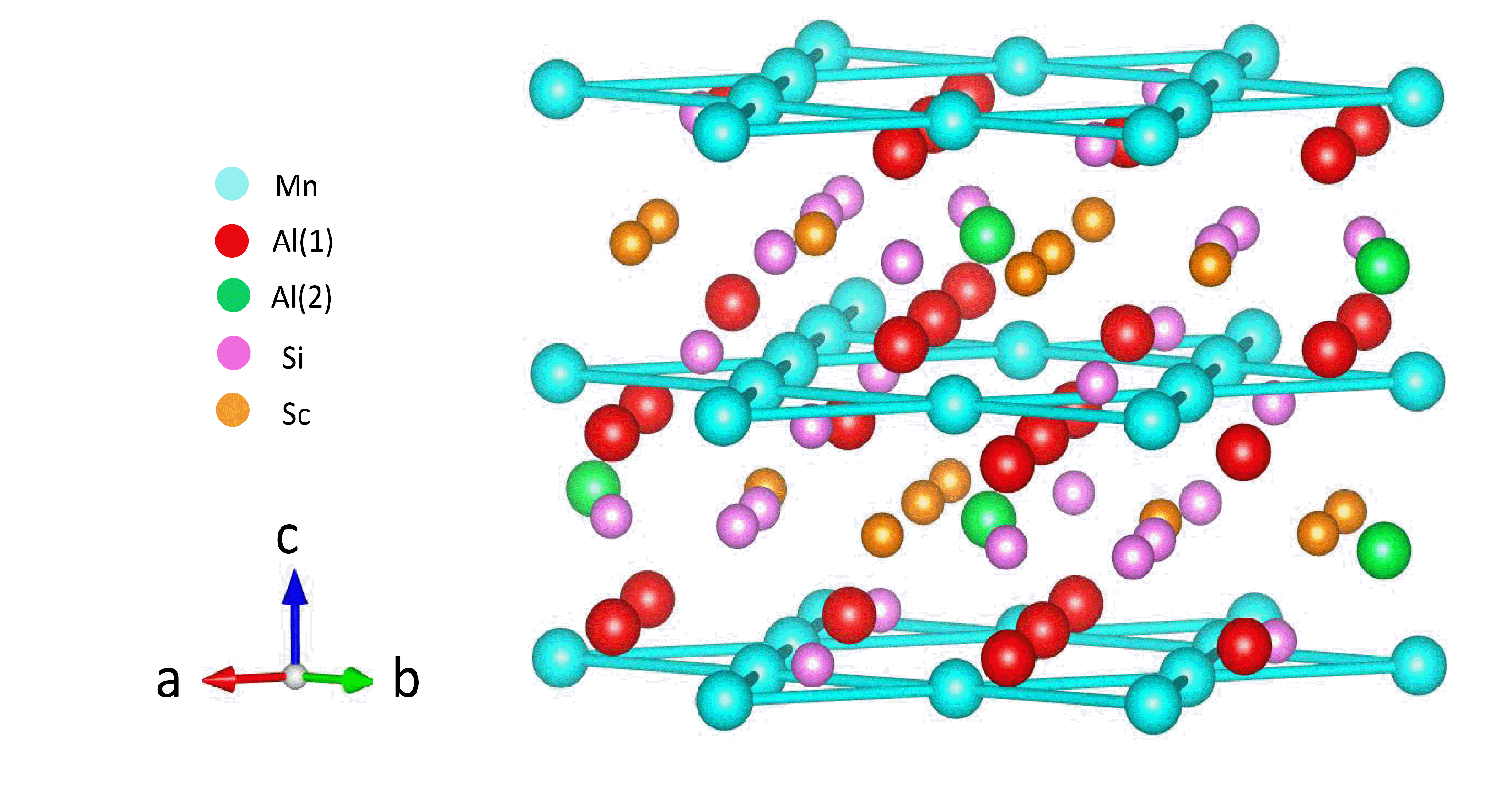} 
\caption{Schematic view of the crystal structure of Sc$_3$Mn$_3$Al$_7$Si$_5$. Note Si atoms are represented by blue circles without showing the two different Si sites, 6h and 4f.}
\label{fig:structure}
\end{figure}

   Recently, particular attention has been paid to the Mn-based kagome metal Sc$_3$Mn$_3$Al$_7$Si$_5$, which has been proposed to lie near a ferromagnetic instability associated with kagome flat-band physics \cite{Samanta2024}. 
   Sc$_3$Mn$_3$Al$_7$Si$_5$ crystallizes in a hexagonal structure (space group: P6$_3/mmc$) with the lattice parameters of $a$ = $b$ = 8.3519(2)~\AA~ and $c$ = 9.0845(4)~\AA~at $T$ = 299 K \cite{He2014}. 
    Its structure consists of two-dimensional Mn kagome layers separated by Sc and Si atoms (Fig.~\ref{fig:structure}) \cite{VESTA}.
 There are six unique crystallographic sites: Sc (6h), Mn (6g), Al1 (12k), Al2 (2b), Si1 (6h), and Si2 (4f) with the numbers in parentheses representing the Wyckoff position of each atom. 
        The metallic nature of the compound has been confirmed by its temperature ($T$) dependence of resistivity. 
     Despite the absence of long-range magnetic order down to 1.8 K \cite{He2014}, several experiments indicate substantial electronic correlations. 
   Specific-heat measurements revealed a large Sommerfeld coefficient of 40--80 mJ/mol K$^2$ \cite{He2014,Li2021,Guehne2026}, while Curie–Weiss analysis of the temperature dependence of magnetic susceptibility yielded large negative Weiss temperatures \cite{He2014,Samanta2024}. 
   Interestingly, Samanta {\it et al}. proposed the existence of ferromagnetic fluctuations based on magnetic susceptibility, $^{27}$Al nuclear magnetic resonance (NMR), transport, and optical-conductivity measurements, despite the predominantly antiferromagnetic Weiss temperatures \cite{Samanta2024}.

 \begin{figure*}[h!tb]
\includegraphics[width=1.8\columnwidth]{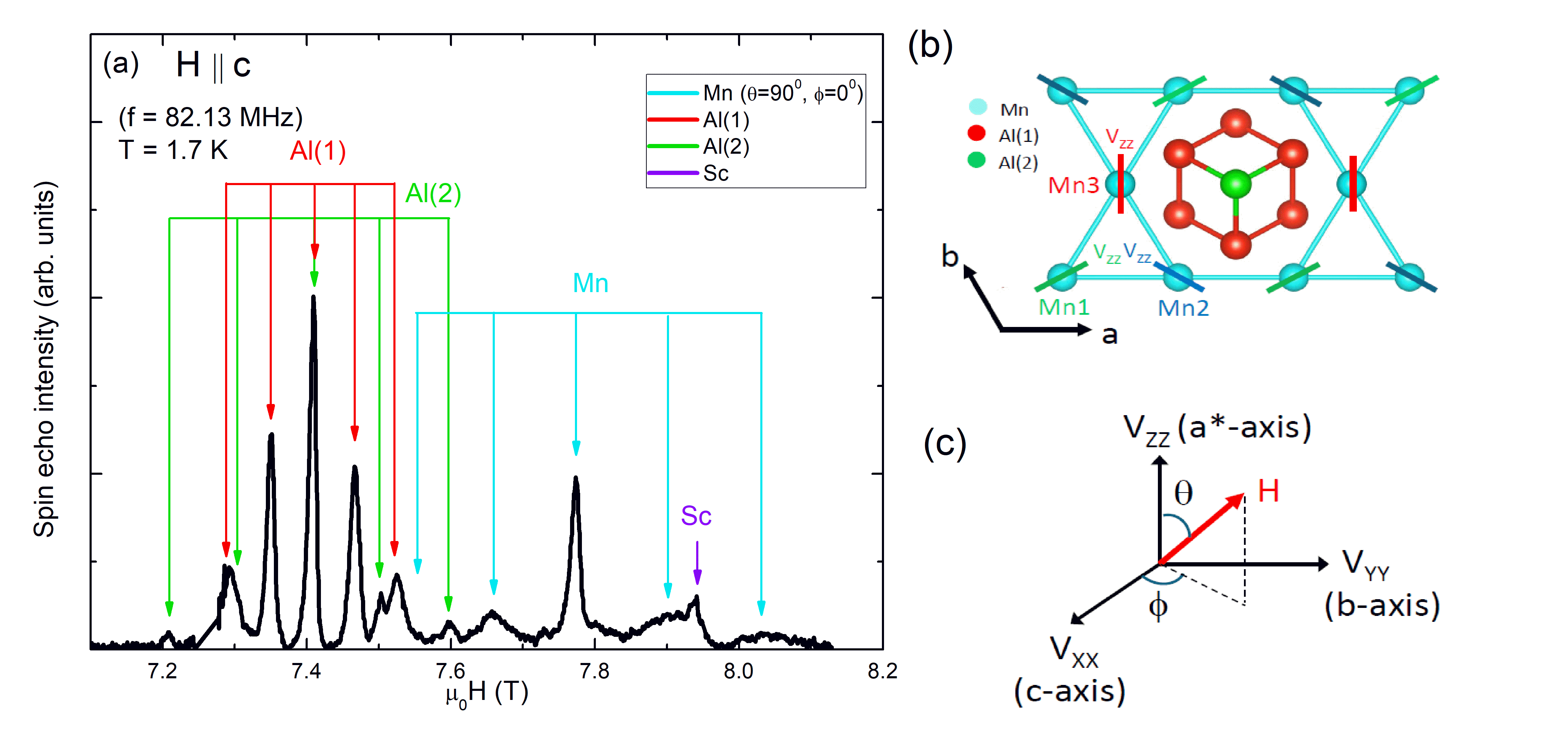}
\caption{(a) $H$-swept $^{55}$Mn-NMR spectrum for $H \parallel c$ measured at 1.7 K and $f$ = 82.13 MHz.  The cyan lines are the positions  $^{55}$Mn-NMR spectra calculated with  $\nu_{\rm Q}$~= 4.6~MHz, $\eta$ = 0.45, $\theta$ = 90$^{\circ}$, and $\phi$ = 0$^{\circ}$.
  The red and green lines are the positions of $^{27}$Al-NMR spectra from Al(1) and Al(2), respectively,  reported previously in Ref. \onlinecite{Ding2025}.  
  The tiny peak around 7.938 T is assigned to $^{45}$Sc NMR signal.  
  (b) Top view of the Mn kagome constructed with Mn ions (cyan) with Al(1) (red) and Al(2) (green).
   The principal axes of $V_{\rm ZZ}$ for three inequivalent Mn sites, Mn1, Mn2 and Mn3, are represented by green, blue, and red short bars having threefold symmetry (i.e., the relative angle between the $V_{\rm ZZ}$ directions for the three Mn sites is 60$^{\circ}$). 
(c) Schematic view of the directions for the principal axes of the EFG with respect to the crystalline axes for Mn1 determined by the NMR measurements and the electronic band structure calculation. $V_{\rm ZZ}$ is parallel to the $a^*$ axis, and   $V_{\rm XX}$ and $V_{\rm YY}$ are nearly parallel to the $c$ and $b$ axes, respectively, for Mn1.  
$\theta$ denotes the polar angle between the $Z$ axis of the EFG and the direction of $H$, and $\phi$ is the azimuthal angle measured from the $V_{\rm XX}$ direction.  }
\label{fig:spectrum_c}
\end{figure*}

Consistently, our previous $^{27}$Al-NMR study on single-crystalline Sc$_3$Mn$_3$Al$_7$Si$_5$ demonstrated that ferromagnetic fluctuations become significant and dominate at low temperatures \cite{Ding2025}. 
On the other hand, a recent $^{55}$Mn-NMR study performed with magnetic field applied along the $c$ axis suggested the emergence of spin-singlet pairing below approximately 10 K \cite{Guehne2026}. 
   Since these interpretations imply markedly different low-energy spin fluctuations, further microscopic investigation is required. 
   In this work, we perform detailed $^{55}$Mn NMR  measurements on single crystals of Sc$_3$Mn$_3$Al$_7$Si$_5$ with magnetic fields applied both parallel and perpendicular to the $c$ axis and $^{55}$Mn nuclear quadrupole resonance (NQR) measurements to clarify the intrinsic low-energy spin dynamics and magnetic correlations in this kagome metal.

     \section{Experimental Details}
   Rod-like single crystals of Sc$_3$Mn$_3$Al$_7$Si$_5$ with hexagonal-shaped cross-section were grown out of high temperature solutions \cite{He2014}.
      The crystalline $c$ axis and the $ab$ plane are parallel and perpendicular to the rod direction of the crystal, respectively. 
 
      NMR/NQR  measurements of $^{55}$Mn ($I$ = $\frac{5}{2}$, $\frac{\gamma_{\rm N}}{2\pi}$ = 10.50 MHz/T, $Q=$ 0.33$\times$10$^{-28}$ m$^2$) nuclei were conducted using a laboratory-built phase-coherent spin-echo pulse spectrometer. 
       The NMR spectra were obtained by sweeping external magnetic field $H$ at fixed frequencies. 
      $H$ was applied parallel to either the crystalline $c$ axis ($H \parallel c$) or the $ab$ plane ($H$ $\parallel$ $ab$).
      The NQR spectrum in zero magnetic field was measured in steps of frequency by plotting the spin-echo intensity.  
      The $^{55}$Mn-NMR spin-lattice relaxation rates 1/$T_{1}$ were measured with a saturation recovery method.
       $1/T_1$ at each $T$ was determined by fitting the nuclear magnetization $M$ versus time $t$  using the exponential function $1-M(t)/M(\infty) = 0.028e^{-t/T_1}+0.178e^{-6t/T_1}+0.794e^{-15t/T_1}$,  where $M(t)$ and $M(\infty)$ are the nuclear magnetization at time $t$ after the saturation and the equilibrium nuclear magnetization at $t$ $\rightarrow$ $\infty$, respectively, for the case of magnetic relaxation \cite{Recovery}. 
     All the nuclear magnetization recovery curves were well fit by the function. 
     Nuclear spin-spin relaxation behavior was inspected by measuring the spin echo intensity ($M_{xy}$) as a function of time ($\tau$) between the first and second pulses with different widths and $rf$ power.

    DFT calculations were performed using a full-potential linear augmented plane wave (FP-LAPW) method, as implemented in \textsc{wien2k}~\cite{FLAPW, Blaha2020}.
Experimental lattice parameters~\cite{He2014} are adopted and the primitive cell contains two formula units.
The generalized gradient approximation of Perdew, Burke, and Ernzerhof~\cite{GGA} was used for the correlation and exchange potentials.
To generate the self-consistent potential and charge, we employed $R_\text{MT}\cdot K_\text{max}=8.0$ with Muffin-tin (MT) radii $R_\text{MT}=$ 2.4, 2.4, 2.1, and 2.1 {a.u.}, for Sc, Mn, Si and Al, respectively. The calculations were performed with 144 $k$-points in the irreducible Brillouin zone (IBZ) and were iterated until charge differences between consecutive iterations are smaller than $1\times10^{-3}$ $e$ and the total energy difference lower than 0.01 mRy.

 \begin{figure*}[h!tb]
\includegraphics[width=1.7\columnwidth]{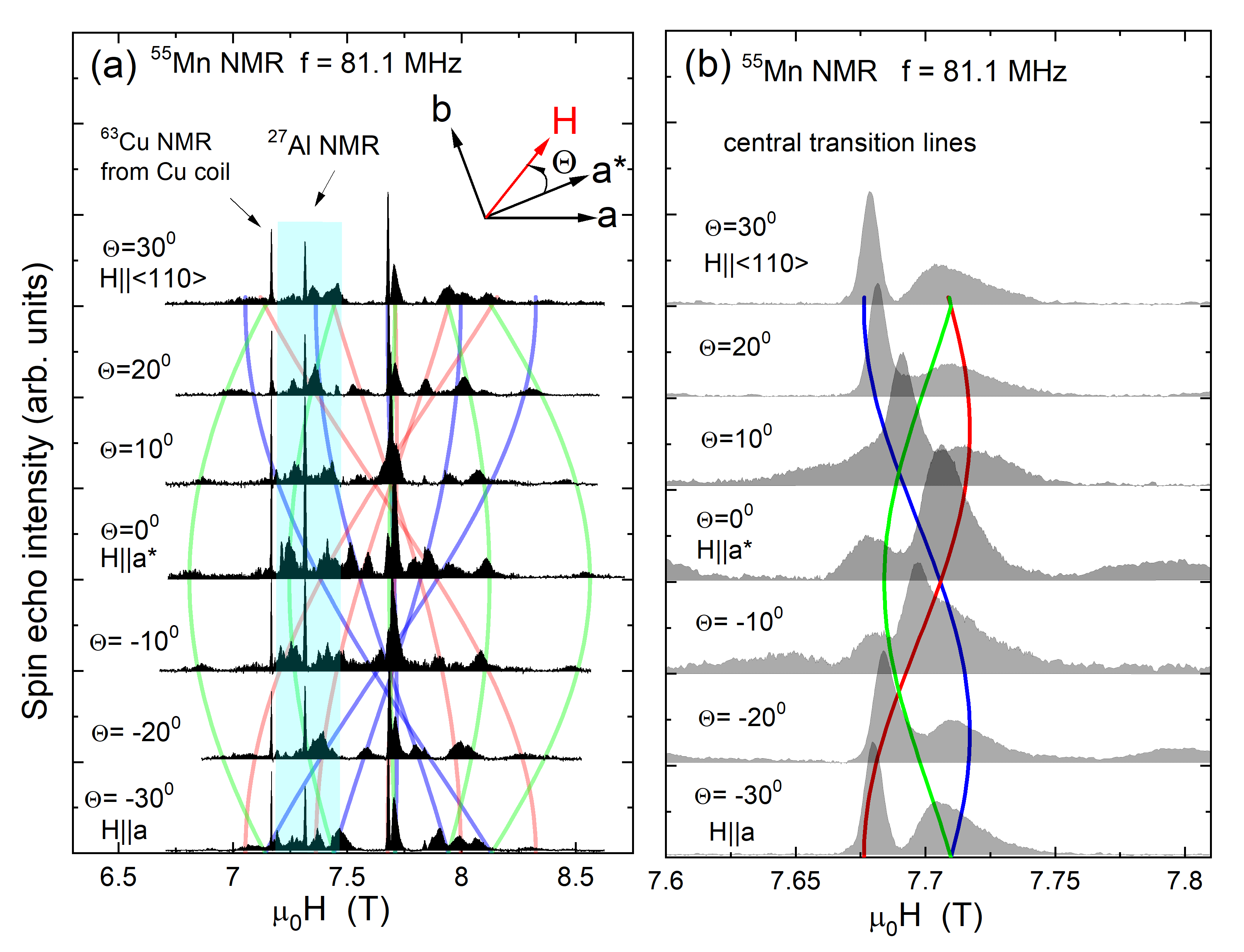}
\caption{(a) Angle ($\Theta$) dependence of $H$-swept $^{55}$Mn-NMR spectra for $H \perp c$ at 4.2 K for the  case of $f$ = 81.1 MHz.  The green, blue,  and red curves represent the calculated $\Theta$ dependence of the peak position of  $^{55}$Mn-NMR spectra using  $\nu_{\rm Q}$ = 4.6 MHz and  $\eta$ = 0.45  for the Mn1, Mn2, and Mn3 sites, respectively.  The inset illustrates a schematic view of the angle $\Theta$.  The peaks in the light blue shaded region in the figure mainly originate from $^{27}$Al NMR signals of the Al(1) and Al(2) sites. 
(b) $\Theta$ dependence of the central transition lines zoomed from panel (a).  The curves show the calculated $\Theta$ dependence of the central transition lines for the Mn1 (green), Mn2 (blue), and Mn3 (red) sites as in panel (a). Here $\Theta$-independent Knight shift $K$ = 0.5 \% is used for the calculation.  }
\label{fig:spectrum_ab}
\end{figure*}

 \section{Results and discussion}
 \subsection{ $^{55}$Mn NMR spectrum }

   Figure  \ \ref{fig:spectrum_c}(a)  shows the $H$-swept $^{55}$Mn-NMR spectrum for $H \parallel c$ axis at 1.7 K,
together with the previously reported $^{27}$Al NMR lines from the Al(1) and Al(2) sites \cite{Ding2025}. 
   The typical spectrum for a nucleus with spin $I=5/2$ with Zeeman and quadrupolar interactions can be described by the following nuclear spin Hamiltonian \cite{Slichter_book} 
\begin{eqnarray}
\centering
{\cal H} &=& -\gamma_{\rm N}\hbar(1+K) {\bf H} \cdot {\bf I} + \frac{h\nu_{\rm Q}}{6}(3I_Z^2-I^2  +\frac{1}{2}\eta(I_+^2 +I_-^2)),
\label{eq:1}
\end{eqnarray} 
where $H$ is external field, $\hbar$ is Planck's constant ($h$) divided by 2$\pi$,  and $K$ represents the Knight shift.  
The first and second terms represent Zeeman and quadrupolar interactions, respectively.  
   The nuclear quadrupole frequency for $I=5/2$ nuclei is given by $\nu_{\rm Q} = 3eQV_{\rm ZZ}/20h$, where $Q$ is the nuclear quadrupole moment and $V_{\rm ZZ}$ is the maximum electric field gradient (EFG) at nuclear sites.
  $\eta$ is the asymmetry parameter of EFG  defined by  $\frac{V_{\rm XX} -V_{\rm YY}}{V_{\rm ZZ}}$ with $|V_{\rm ZZ}|$$\geq$$|V_{\rm YY}|$$\geq$$|V_{\rm XX}|$ in the EFG (XYZ) coordinate.

    When the Zeeman interaction is much greater than the quadrupolar interaction, the spectrum consists of a central transition line with two satellite lines on both sides.
    In this case, the resonance frequency $f$ for the transition from $I_z$ = $m$ to $m-1$ is given within first-order perturbation theory by \cite{Abragambook}
 \begin{eqnarray*}
 \lefteqn{ f(m \leftrightarrow m-1)}  \\
& =&   f_0 + \frac{1}{2}\nu_{\rm Q}(m-\frac{1}{2})(3\cos^2\theta-1+\eta \sin^2\theta \cos2\phi ). 
  \label{eq:3}
  \end{eqnarray*} 
       Here $ f_0 = \frac{\gamma_{\rm N}}{2 \pi}H(1+K)$, and $\theta$ and $\phi$ are the polar and azimuthal angles between the $Z$ axis of the EFG and the direction of $H$, respectively, where the quantization axis ($z$ axis) for the Zeeman interaction points along the $H$ direction. 
    Thus, the NMR spectra, especially the spacing between the lines, depend on the angles $\theta$ and $\phi$ as well as $\nu_{\rm Q}$ and $\eta$. 
It is noted that, when ($\theta$, $\phi$) = (0$^{\circ}$, 0$^{\circ}$) (i.e., $H \parallel$  Z axis), the spacing between the outermost satellites $\Delta f$ = $|f(5/2 \leftrightarrow 3/2) - f(-3/2 \leftrightarrow -5/2)|$ is maximum, equal to  4$\nu_{\rm Q}$. 
On the other hand, in the case of ($\theta$, $\phi$)  =  (90$^{\circ}$, 0$^{\circ}$) (i.e., $H \parallel$  X axis),  one has a value of  $|\Delta f|$ = 2$\nu_{\rm Q}(1-\eta$). 
When $H$ is parallel to the Y axis [i.e.,  ($\theta$, $\phi$) = (90$^{\circ}$, 90$^{\circ}$)],   $|\Delta f|$ = 2$\nu_{\rm Q}$(1+$\eta$).

To obtain information about the values of $\eta$ and $\nu_{\rm Q}$, and the principal axes of the EFG at the Mn site,  we have calculated the EFG at the Mn site by using DFT calculations. 
    We found from the calculations that  $\nu_{\rm Q}$ = 4.60 MHz and the $\eta$ = 0.26 which are different from the DFT calculation results of $\nu_{\rm Q}$ = 5.266 MHz and the $\eta$ = 0 reported by Guehne {\it et al} \cite{Guehne2026}.  
    It is noted that, since the local symmetry of the Mn site is .2/$m$.  \cite{He2014}, $\eta$ should not be zero.
     We also found that a crystallographically unique Mn site has three different local EFG coordinates due to the local symmetry as shown in Fig. \ \ref{fig:spectrum_c}(b).  
     For Mn1,  for example, $V_{\rm ZZ}$ is parallel to the $a^*$ axis in the $ab$ plane and the principal axis of $V_{\rm XX}$ is nearly parallel to the $c$ axis as schematically shown in Fig. \ \ref{fig:spectrum_c}(c). 
     As a consequence, when $H$ is applied in the $ab$ plane, three sets of Mn NMR lines will be observed, in contrast to the case of $H$ $||$ $c$ where the three different Mn sites will be equivalent.
   In fact, we observed only a set of five quadrupolar-split NMR lines under $H$ $||$ $c$ as shown in Fig. \ \ref{fig:spectrum_c}(a).  
   
   In contrast, we observed more complicated $^{55}$Mn NMR lines when $H$ is applied in the $ab$ plane. 
   Figure  \ref{fig:spectrum_ab}(a) shows the angle $\Theta$ dependence of $^{55}$Mn NMR spectra measured at $T$ = 4.2 K. Here $\Theta$ is the angle measured from the $a^*$ axis: $\Theta$  = 30$^{\circ}$ and  $-$30$^{\circ}$ are parallel to $<$110$>$ and the $a$ axis, respectively [see the inset of Fig. \ref{fig:spectrum_ab}(a)]. 
    Note that the peaks in the light blue hatched region in the figure mainly originate from $^{27}$Al NMR signals of Al(1) and Al(2).  
   
    It is clear that the spacing between the most outer satellite for one of the Mn sites (that is, Mn1 site) produces a maximum value of 1.75(4) T (= 18.4(5) MHz) at $\Theta$ = 0$^{\circ}$ corresponding $H$ $||$ $a^*$. 
     This yields $\nu_{\rm Q}$ = 4.6(1) MHz.  
     Utilizing the $\nu_{\rm Q}$ = 4.6 MHz, the $^{55}$Mn NMR spectrum for $H||c$ [with ($\theta$, $\phi$)  =  (90$^{\circ}$, 0$^{\circ}$)] is also reasonably reproduced as shown by light blue lines in Fig. \ref{fig:spectrum_c}(a) where we used $\eta$ = 0.45(2). 
     These parameters also consistently reproduce the $\Theta$  dependence of $^{55}$Mn NMR spectra under $H||ab$ plane as shown in Fig. \ref{fig:spectrum_ab}(a) where NMR lines from three inequivalent Mn sites are observed. 
     The green, blue, and red curves are the calculated angular dependence of the Mn-NMR line positions for the Mn1, Mn2, and Mn3 sites, respectively.   
     Here we calculated the angle dependence of the peak positions for three Mn sites by fully diagonalizing the nuclear spin Hamiltonian [Eq. (\ref{eq:1})] to ensure that the effects of quadrupolar interaction and Knight shift were fully accounted for in the determination of the positions of NMR lines.
    The angular dependence of the central transition lines is also well reproduced by the calculation [Fig. \ref{fig:spectrum_ab}(b)], where an angle-independent Knight shift ($K_{ab} = 0.5 \%$) in the $ab$ plane was used.   

    The values of $\nu_{\rm Q}$ and $\eta$ for the Mn site were also determined to be $\nu_{\rm Q}$ = 4.8(1) MHz and $\eta$ = 0.48(2) by the $^{55}$Mn nuclear quadrupole resonance (NQR) spectrum shown below, whose values are close to the values determined by the $^{55}$ NMR-spectrum measurements described above.

 \subsection{ $^{55}$Mn Knight shift}

 \begin{figure}[tb]
\includegraphics[width=1\columnwidth]{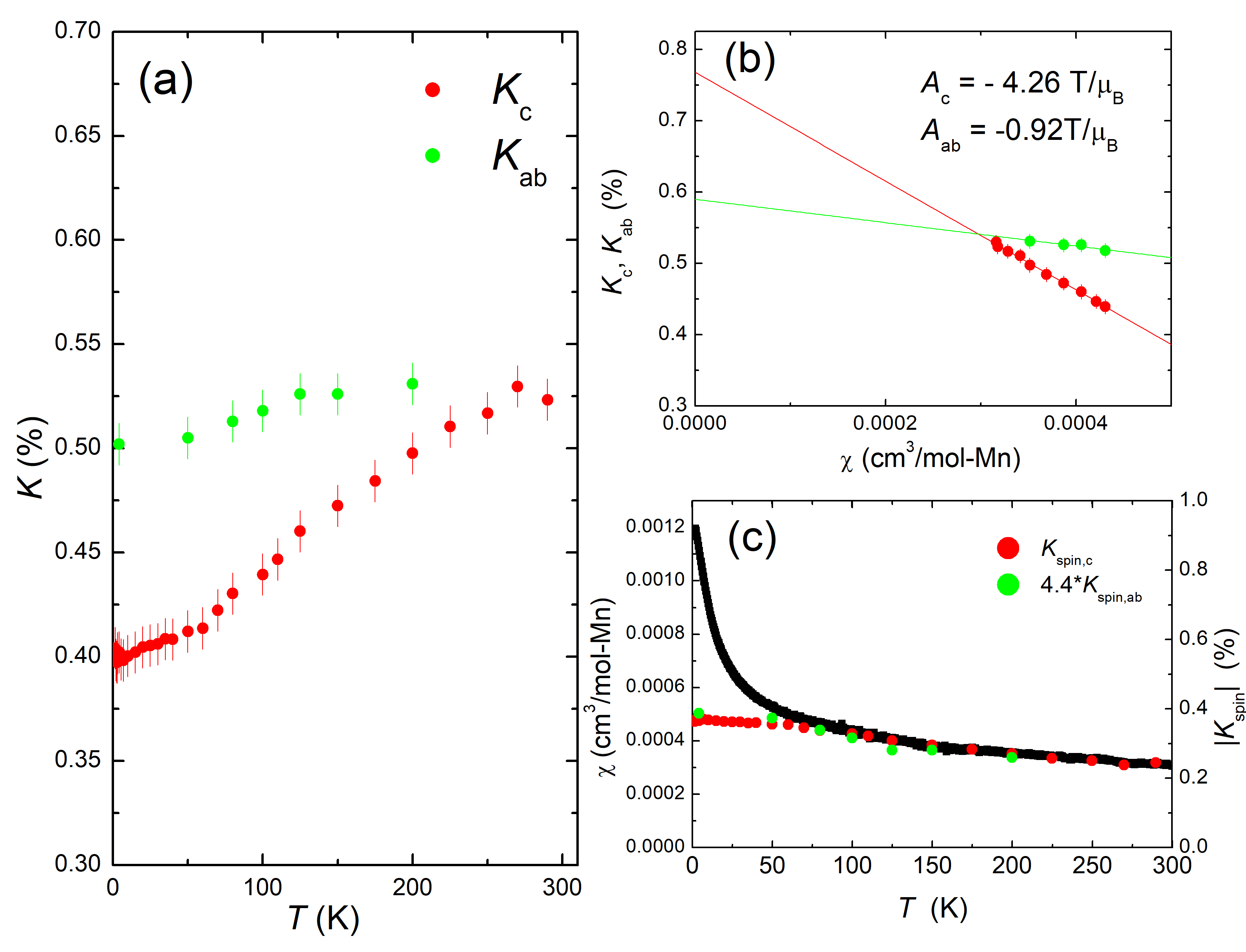} 
\caption{(a) $T$ dependence of Knight shifts $K_c$  and $K_{ab}$. (b) $K$-$\chi$ plot for $K_c$ and $K_{ab}$ versus the corresponding $\chi$. (c) $T$ dependence of the spin part of the Knight shift, $|K_{\rm spin,c}|$  and $|K_{\rm spin,ab}|$,   together with the $T$ dependence of $\chi_{c}$ (black circles) measured at 7.4 T \cite{Chi}.  Note that  $|K_{\rm spin,ab}|$ values were multiplied by a factor of 4.4 to scale with  $|K_{\rm spin,c}|$ due to the different hyperfine coupling constants. }
\label{fig:Shift}
\end{figure}

  Figure \ \ref{fig:Shift}(a) shows the $T$ dependence of the Knight shifts for $H\parallel c$ ($K_c$)  and $H \parallel ab$ plane ($K_{ab}$)  determined by fitting the observed spectra. 
    Here we plotted $K_{ab}$ determined by the central line for Mn3 under  $H$ $\parallel$ $a$. 
    As  mentioned above,  we did not see the clear $ab$-plane anisotropy in Knight shift within our experimental uncertainty. 
   Both $K_c$  and $K_{ab}$ decrease slightly with decreasing $T$ from 300 K to $\sim$ 50 K and level off at lower $T$.
     A similar temperature dependence has been reported in  $^{27}$Al Knight shift \cite{Ding2025} and  $^{55}$Mn Knight shift  \cite{Guehne2026} measurements in Sc$_3$Mn$_3$Al$_7$Si$_5$. 

      The Knight shift has contributions from the $T$-dependent spin part $K_{\rm spin}$ and a $T$-independent orbital part $K_0$. 
      $K_{\rm spin}$ is proportional to the spin susceptibility $\chi_{\rm spin}$ through the hyperfine coupling constant $A$ giving $K=K_0+\frac{A}{N_{\rm A}\mu_{\rm B}}\chi_{\rm spin}$, where $N_{\rm A}$ is Avogadro's number.
     Figure \ref{fig:Shift}(b) plots $K_{ab}$ and $K_{c}$ against the corresponding  $\chi_{ab}$  and  $\chi_c$, respectively, with $T$ as an implicit parameter. 
     Here we used the $K_{ab}$ and $K_{c}$ data above 100 K to avoid the component of the magnetic susceptibility data originating from the extrinsic contribution observed at low $T$. 
     As shown,  $K_{ab}$ and $K_{c}$ are seen to vary nearly linearly with the corresponding $\chi$, and the hyperfine coupling constants are estimated to be $A_{c}$ = --4.26(2) T/$\mu_{\rm B}$, and $A_{ab}$ = --0.92(5) T/$\mu_{\rm B}$ for $H$ $\parallel$ $c$  and $H$ $\parallel$ $ab$, respectively, from the slopes of the linear lines.
  These values are smaller than a typical value $A$ = -10.5~T/$\mu_{\rm B}$ for 3$d$ electron core polarization \cite{Abragam1955}.
      Similar reductions in the hyperfine coupling constant of 3$d$ elements have been observed in several other compounds \cite{Tsuda1968,Fukai1996,Roy2013,PaulPRB}.
    As discussed in Refs. \onlinecite{Tsuda1968,Fukai1996,Roy2013}, the small values could be due to anisotropic and positive orbital and/or dipolar hyperfine coupling contributions which cancel a part of the negative core polarization hyperfine field.    
    The values of $K_0$ for $H$ $||$ $c$ and  $H$ $\parallel$ $ab$ are estimated to 0.77, and  0.59 \%, respectively.

  Utilizing the $K_0$ values, the spin part of the Knight shifts for each direction was evaluated. 
   Figure \ \ref{fig:Shift}(c) exhibits the temperature dependence of the absolute values of $|K_{\rm spin,c}|$  and $|K_{\rm spin,ab}|$ together with the $T$ dependence of $\chi_{c}$ (black circles) measured at 7.4 T \cite{Chi}.  
 Here $|K_{\rm spin,ab}|$ values were multiplied by a factor of 4.4 to scale with  $|K_{\rm spin,c}|$ due to the different hyperfine coupling constants.
 The $T$ dependence of $|K_{\rm spin,c}|$  and $|K_{\rm spin,ab}|$ scales well with $\chi$ above $\sim 100$ K where $\chi_{c}$ was tentatively plotted as $\chi$ is isotropic \cite{Ding2025}. 
   In contrast, a clear deviation between $K_{\rm spin} $ and $\chi$  can be seen below $\sim$ 80 K.  
   These results clearly evidenced that the upturn in $\chi$ observed at low $T$ in Fig. \ref{fig:Shift} is non-intrinsic and likely arises from a small number of paramagnetic impurities, disorder, or defects, as have been pointed out by our previous $^{27}$Al-NMR measurements \cite{Ding2025}.     

It is important to point out that neither $|K_{\rm spin,c}|$  and $|K_{\rm spin,ab}|$ in the $^{55}$Mn NMR measurements show any appreciable reduction at low temperatures below 10 K. 
In addition, we did not observe any obvious change in the line width of the $^{55}$Mn NMR lines at low temperatures. 

  \subsection{ $^{55}$Mn spin-lattice relaxation time $T_1$ and spin-spin relaxation time $T_2$}
 
\begin{figure}[tb]
\includegraphics[width=0.9\columnwidth]{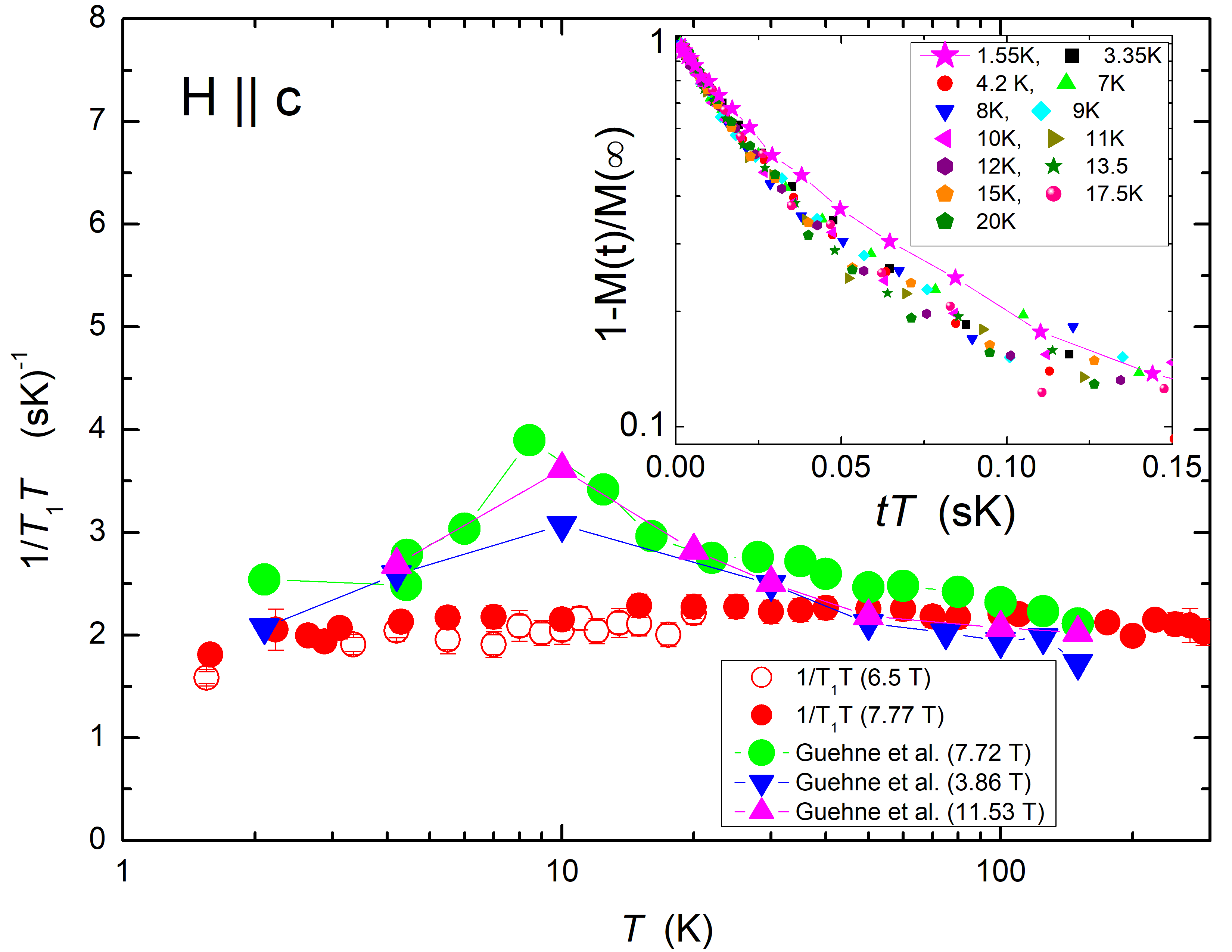} 
\caption{(a) $T$ dependence of $^{55}$Mn-1/$T_1T$ measured at 6.5 T and 7.77 T under $H$ $||$ $c$. $^{55}$Mn-1/$T_1T$ data reported in Ref. \cite{Guehne2026} are also plotted. 
Inset shows  the nuclear magnetization recovery curves at several temperatures below 20 K  plotted as a function of $tT$, the time $t$ multiplied by temperature $T$. }
\label{fig:T1}
\end{figure}

 Figure~\ref{fig:T1} shows the temperature dependence of 1/$T_1T$ measured at the $^{55}$Mn-NMR central transition line under $H$ $||$ $c$.   
 $1/T_1T$ is nearly constant over the entire $T$ range, evidencing the metallic state from a microscopic point of view. 
   No enhancement or peak is observed around 10 K, in contrast to the results reported in Ref.~\cite{Guehne2026}, whose data are also included in the figure for comparison.
 
The inset shows the nuclear magnetization recovery curves at several temperatures below 20 K where the horizontal axis is $tT$, the time $t$ multiplied by temperature $T$.  As shown, all recovery curves, except for 1.55 K,  are scaled within experimental uncertainty.
This further confirms the $T_1T$ = constant behavior above $\sim$ 3.35 K, showing no enhancement of 1/$T_1T$ around 10 K.   
   Together with the absence of any reduction in $|K_{\rm spin}|$ and line width below 10 K, our $^{55}$Mn NMR results do not show any clear  signature of spin-singlet pairing in  Sc$_3$Mn$_3$Al$_7$Si$_5$. 
    At present, the origin of the discrepancy with Ref.~\cite{Guehne2026} is not yet clear. 
   One possible explanation is sample-dependent effects.
 Although the temperature dependence of the resistivity of our crystal \cite{Nakajima_unpublish} is very similar to the reported ones in Refs. \cite{Samanta2024, He2014,Guehne2026},  a subtle anomaly near $T\sim 12$ K has been reported in the specific heat of the crystal studied in Ref.~\cite{Guehne2026}, which was not observed in the previous study \cite{Samanta2024} or in our sample \cite{Nakajima_unpublish}.
      It is noted that the recovery curve at 1.55 K is slightly shifted as shown in the inset of Fig. ~\ref{fig:T1}. 
   This suggests a tendency for 1/$T_1T$ to decrease slightly below 3.35 K, that is, a slight decrease in the density of states at the Fermi energy. 
   To investigate the behavior in detail, it is necessary to measure $1/T_1T$ at low temperatures below 1.5 K, which is left for future work.

   \begin{figure*}[tb]
\includegraphics[width=2.0\columnwidth]{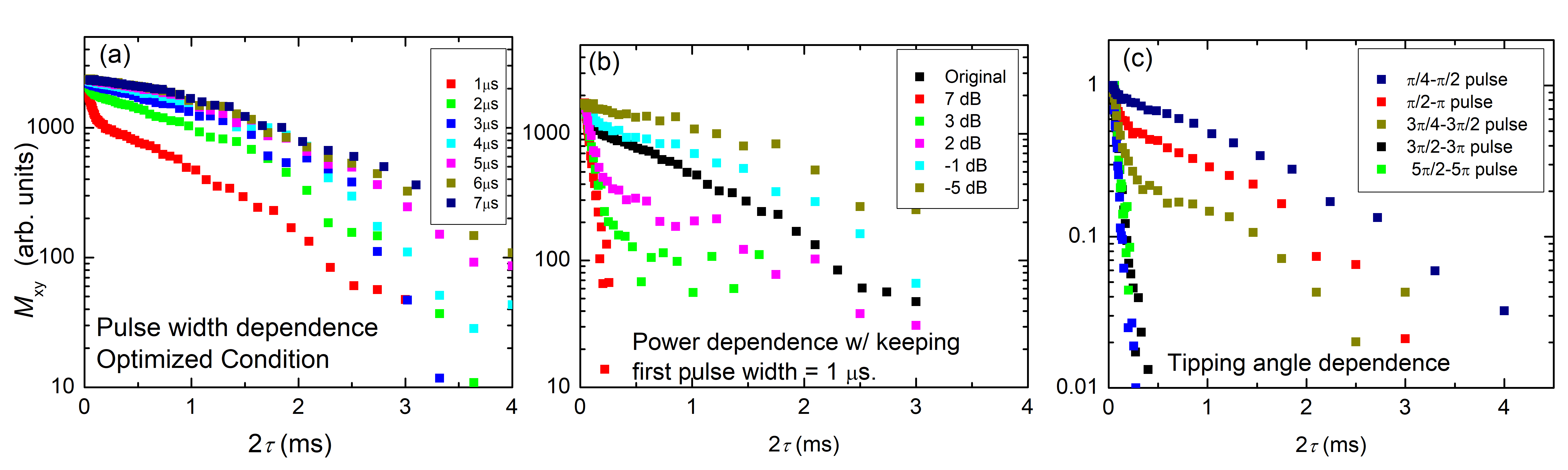} 
\caption{$^{55}$Mn NMR transverse magnetization ($M_{xy}$) decay curves as a function of 2 $\tau$ measured at $T$ = 4.2 K under $H$ $||$ $c$ measured at various conditions. (a) Pulse width dependence. Here $rf$ power was adjusted to optimize the spin-echo intensity (i.e., Hahn echo sequence of $\pi$/2-$\pi$ pulses). The legend indicates the widths for the first ($\pi$/2) pulse. (b) $rf$ power dependence of nuclear magnetization decay curves measured with a fixed first pulse width of 1$\mu$s. The powers described here refer to the power for the Hahn echo condition.  (c) Nuclear spin rotational angle dependence. The pulse widths were changed while keeping the $rf$ power constant to make different nuclear spin rotation sequences from $\pi$/2-$\pi$ to e.g. $3\pi$/2-$3\pi$ and  $5\pi$/2-$5\pi$.    }
\label{fig:T2}
\end{figure*}
  
  Finally we comment on the nuclear magnetization decay curves measured at different conditions. 
  Pulse-width dependence of the nuclear decay curves has been reported by Guehne {\it et al}.  \cite{Guehne2026} where, by increasing pulse width, a Lorentzian decay changes to a Gaussian one. 
   This behavior was explained in terms of the different numbers of nuclei excited by different pulse widths where a change in nuclear-nuclear dipolar couplings from strong indirect to direct has been pointed out for the different decay profiles  \cite{Guehne2026}.  
   To examine this effect, we also measured the nuclear magnetization decay curves with different pulse widths and also its $rf$ power at the central transition line of $^{55}$Mn NMR under $H$ $||$ $c$ at 4.2 K. 
    Figure  \ref{fig:T2} (a) shows the nuclear decay behavior with different pulse widths where the $rf$ power was adjusted to optimize the spin-echo intensity (i.e., Hahn echo sequence of $\pi$/2-$\pi$ pulses). 
   For a $\pi$/2 pulse width of 1 $\mu$s, we observed  two components:  short Lorentzian and long Gaussian decay components. 
     With increasing the pulse width, the Lorentzian component is progressively suppressed, and only the Gaussian component can be observed with larger pulse width.  
    Since the pulse width ($t_{\rm w}$) is inversely related to the excitation frequency width, this result suggests that the reducing the number of excited  nuclei by the $rf$ pulse changes the decay profile from the Lorentzian to Gaussian. This is consistent with the results in  Ref. \cite{Guehne2026}. 
    We further investigated  the decay behavior by changing the $rf$ power while keeping the first (1$\mu$s) and second (2$\mu$s) pulse width as shown in Fig.  \ref{fig:T2} (b). 
    Here we show the different $rf$ powers in dB relative to the optimal condition satisfying the Hahn echo condition ($\pi$/2-$\pi$ pulses). 
    The $rf$ power relates to the effective $rf$ field so-called $H_1$ in NMR experiments, which determines how fast one rotates nuclear spins. 
   It is noted that  the rotational angle ($\theta_{\rm N}$) of nuclear spins is determined by $\theta_{\rm N}$ = $\gamma_{\rm N}$$H_1$$t_{\rm w}$. 
    As can be seen, when the $\theta_{\rm N}$ is large (that is, +7 dB for the power), we observed only the Lorentzian component. 
    On the other hand, only the Gaussian component can be detected when the $rf$ power is reduced. 
Since the reduction of the $rf$ power reduces the number of nuclei excited by the $rf$ pulses, this result would be consistent with the pulse width dependence of decay profiles shown in  Fig.  \ref{fig:T2} (a).
     
     It is also interesting to show the change in the decay profile by changing the nuclear excitations in a different way, where we changed the pulse width while keeping the $rf$ power constant [Fig.  \ref{fig:T2} (c)].
    Here we first optimized the $rf$ power with the first pulse width (1 $\mu$s for the $\pi$/2 pulse) to maximize the spin echo intensity, and we then changed  the pulse width sequences from $\pi$/2-$\pi$ to $3\pi$/2-$3\pi$ and  $5\pi$/2-$5\pi$. 
    In this case, the Gaussian decay component disappears and only the short Lorentzian decay component can be observed with longer pulse widths. 
    Since longer pulse widths reduce the excitation frequency width and thus the number of excited nuclei, it looks like that the behavior is inconsistent with the explanation given above where the reduction in the number of nuclei induces the Gaussian behavior. 
     At present we do not have a clear explanation for the observation. 
     However, one should also consider not only intrinsics but also possible extrinsic effects, such as $H_1$ inhomogeneity arising from the skin depth of the metallic Sc$_3$Mn$_3$Al$_7$S$_5$ and the geometry of the NMR coils.  
     These factors can produce a Gaussian distribution in the nuclear spin rotational angular frequency, ultimately leading to a Gaussian decay behavior of the nuclear magnetization.
   We note that these effects may not originate from the heat-up of the crystal due to eddy current produced by the $rf$ pulses, as we observed no change in the decay profile by changing the repetition time from 50 ms to 5 s.
   It is also important to note that $T_1$ values measured with the spin echoes showing the Lorentzian and Gaussian decays are the same within our experimental uncertainty, again indicating no heat-up issues in the experiments.
   
 \begin{figure}[tb]
\includegraphics[width=0.9\columnwidth]{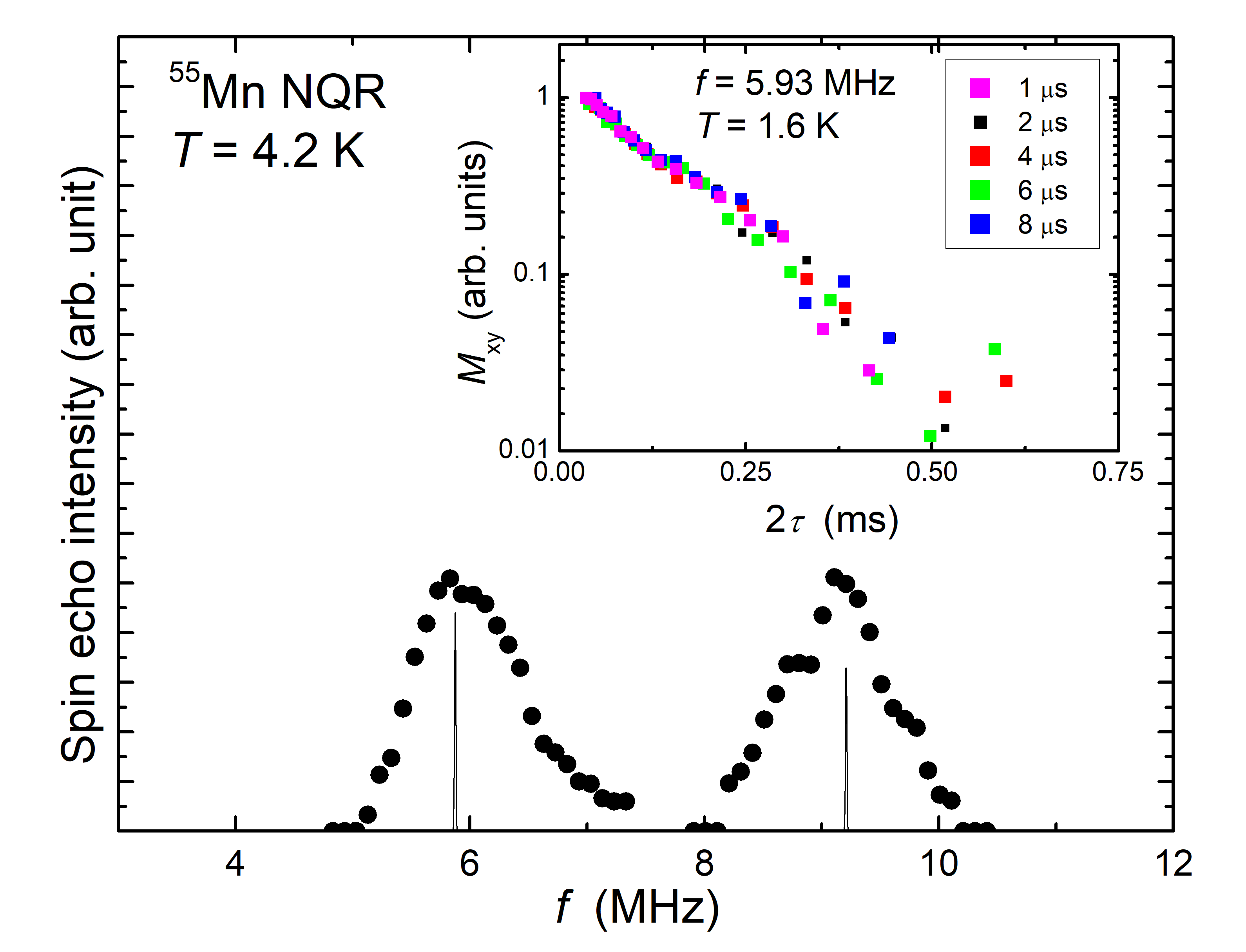} 
\caption{$^{55}$Mn-NQR spectrum observed at $T$ = 4.2 K under zero magnetic field. The two vertical lines represent the calculated NQR line positions for the case of $\nu_{\rm Q}$ = 4.8 MHz and $\eta$ = 0.48. The inset shows the pulse width dependence of $^{55}$Mn NQR transverse magnetization ($M_{xy}$) decay curves as a function of 2 $\tau$ measured at $T$ = 1.6 K and 5.93 MHz using single crystals. The legend indicates the widths for the first ($\pi$/2) pulse.}
\label{fig:NQR}
\end{figure}

   To obtain further insights into the unusual nuclear magnetization decay behaviors observed in the $^{55}$Mn-NMR measurements, we also measured the pulse-width dependence of the decay profiles as in the case of Fig. \ref{fig:T2}(a) using $^{55}$Mn-NQR lines shown in Fig. \ref{fig:NQR}. 
   As shown in the inset of Fig. \ref{fig:NQR}, no clear pulse-width dependence in nuclear decay behavior was observed, where a Lorentzian component was dominant. 
   Furthermore, no difference in the decay behavior between the single crystal and powder samples was observed (not shown here). 
   These results suggest that the Gaussian behavior in the nuclear decay curves observed in the $^{55}$Mn-NMR measurements seems to be a feature specific to the NMR lines, not to the NQR lines.  
   Again, we do not have a clear idea to explain the behavior.    
   Further detailed experimental investigations as well as theoretical studies such as in Ref. \cite{Curr2023} are required to fully understand the origin of the observed decay behavior, which is beyond the scope of this study.

  \section{Summary}
  In conclusion, we conducted $^{55}$Mn-NMR/NQR measurements on  Sc$_3$Mn$_3$Al$_7$Si$_5$ to investigate its static and dynamic magnetic properties. 
  From the detailed NMR/NQR spectrum measurements and the DFT calculations,  we determined the electric field gradient parameters at the Mn site, including a quadrupolar frequency $\nu_{\rm Q}$ = 4.6$\sim$4.8 MHz and an asymmetry parameter $\eta$ = 0.45$\sim$0.48. 
    From the temperature dependence of 1/$T_1T$, Knight shift and line width of  $^{55}$Mn-NMR,  no clear signature of the formation of spin singlet state was observed around 10 K in our Sc$_3$Mn$_3$Al$_7$Si$_5$ crystals, although a slight decrease in 1/$T_1T$ was observed below  $\sim$3-4 K.

 \section{Acknowledgments}
  Y.F. acknowledges Dr. V. Ogloblichev and Prof. A. Mahajan for valuable discussions.  Y.F. also acknowledges Prof. Y. Ihara for fruitful discussions and for showing his NMR results prior to the publication of the data. The research was supported by the U.S. Department of Energy, Office of Basic Energy Sciences, Division of Materials Sciences and Engineering. Ames National Laboratory is operated for the U.S. Department of Energy by Iowa State University under Contract No.~DE-AC02-07CH11358. C.D. and Y.N. were supported by an NSF Career DMR-1944975.


\end{document}